\documentclass[conference]{IEEEtran}
\IEEEoverridecommandlockouts

\usepackage{cite}
\usepackage{amsmath,amssymb,amsfonts}
\usepackage{graphicx}
\usepackage{subcaption}
\usepackage{textcomp}
\usepackage{xcolor}
\usepackage{booktabs}
\usepackage{multirow}
\usepackage{url}
\usepackage{balance}
\usepackage{microtype}
\usepackage{comment}

\begin{document}

\title{Multi-Class, Multi-Tier Network Intrusion Detection: A Comprehensive and Reproducible Benchmark}


\author{
  \IEEEauthorblockN{Yufeng Xin and Bryant Goseland}
  \IEEEauthorblockA{RENCI, University of North Carolina at Chapel Hill\\
    Chapel Hill, NC, USA\\
    yxin,goselandb@renci.org}
  \and
  \IEEEauthorblockN{Mohamed Rahouti}
  \IEEEauthorblockA{Dept. of Computer and Information Science \\
  Fordham University \\
  New York, NY, USA \\
  mrahouti@fordham.edu}
}

\maketitle

\begin{abstract}
Machine learning (ML) and deep learning (DL) have dominated Intrusion Detection System (IDS) research in recent years. Unfortunately, many existing studies have produced inflated results and unreliable benchmarks due to critical oversights and mistakes in the ML and DL pipeline, from data collection and labeling to feature engineering and model training and evaluation. CIC-IDS2017 is a standard benchmark for network intrusion detection. Still, many published results on this dataset are difficult to compare due to labeling errors, inconsistent flow extraction, potential leakage, and performance evaluation metrics dominated by benign traffic.
In this paper, we present a comprehensive benchmark with corrected PCAP-level labeling and a complete evaluation pipeline with diverse ML models. We evaluate eleven tabular classifiers at three nested levels: binary attack detection, nine-class attack-family attribution, and fifteen-class fine-grained classification.
A soft-voting ensemble of Random Forest, XGBoost, and LightGBM obtains the best fine-tier macro-F1 of 0.955, with coarse and binary macro-F1 scores of 0.980 and 0.999, respectively. We further conducted a feature selection study based on an analysis of feature importance. This comprehensive benchmark pipeline is configurable and open-source, enabling new feature extraction and model plugins for new datasets. Future work should use this pipeline as a reference point for richer features, rare-class analysis, and model generalization towards new datasets and attack classes.
\end{abstract}


\section{Introduction}
\label{sec:intro}

Network intrusion detection systems (NIDS) increasingly use machine-learning classifiers trained on flow-level traffic summaries.
This trend is especially relevant to future networks, where 5G/6G, edge, IoT, and software-defined infrastructures are expected to support automated security monitoring and rapid response at scale.
For these systems, benchmark construction is as important as model choice: label noise, leakage, and majority-class-dominated metrics help avoid common ML pitfalls~\cite{lones2024avoiding}.
CIC-IDS2017~\cite{sharafaldin2018cicids} is among the most widely used public NIDS datasets because it provides five days of enterprise-like traffic, raw PCAPs, and a broad set of attack scenarios~\cite{panwar2022intrusion,liu2024elephant,ahmed2024explainable,sinha2025efficient,bijlwan2026deep,hadi2026genti}.
Its popularity, however, has also propagated weaknesses in the benchmark.
Prior research report substantial label corruption, including per-attack corruption above 90\% in some web-attack settings~\cite{liu2022errorprevalence}, and a recent study documents discrepancies 
between the public attack description and the packets visible in the captured subnet~\cite{lycos2017}. Many recent studies also based their model performance claims on the easier binary classification problem.

These issues make many reported CIC-IDS2017 results difficult to interpret and hard to reproduce. Making the problem worse is the misuse of evaluation metrics, especially studies that emphasize accuracy or weighted F1 above $99\%$.
Those aggregate metrics are dominated by benign traffic and do not show whether a model distinguishes attacks at finer granularity, esp. the minority attacks such 
as Bot, Web Attack, Heartbleed, or SQL Injection that appear in only a very small number of flows in the dataset.
Results based on uncorrected CICFlowMeter CSV files, testbed-specific metadata, or binary-only evaluation also do not reliably establish attack-family or fine-grained class attribution.
A credible benchmark, therefore, requires corrected labels derived from the PCAPs, a common preprocessing and splitting protocol that excludes metadata leakage, and metrics reported at multiple class granularities.

This paper presents a comprehensive benchmark and a complete multi-class, multi-tier IDS evaluation pipeline with three major contributions:
\begin{enumerate}
  \item It defines a reproducible PCAP-level labeling pipeline for CIC-IDS2017 that corrects NAT address translation, web-attack filtering, Heartbleed source-port ambiguity, and Infiltration label precision through exact 5-tuple flow-key matching.
  \item It compares eleven tabular classifiers under the same feature extraction, preprocessing, and train/validation/test protocol, covering linear models, bagging, boosting, neural tabular models, and ensemble meta-learners.
  \item It reports a three-tier evaluation---binary, nine-class coarse, and fifteen-class fine---using macro-F1 as the primary metric so that minority-class failures are visible rather than hidden by weighted averages.
  \item It provides a feature importance analysis that helps enhance model training and interpretability.
\end{enumerate}

The benchmark shows that gradient-boosted tree models and a simple soft-voting ensemble are consistently strongest. Binary and coarse-family detection are nearly perfect, and the remaining fine-class errors are concentrated in rare attacks with limited support. This study suggests that future research should not be limited to another tabular model comparison on a similar dataset alone unless it introduces new data, richer features, a rare-class methodology, or cross-dataset validation.
All code, data-processing scripts, and experimental artifacts are available for reproducibility and extension at~\cite{xin2026repo}.

\section{Related Work}
\label{sec:related}

\subsection{CIC-IDS2017 and Label Quality}

Sharafaldin et al.~\cite{sharafaldin2018cicids} introduced CIC-IDS2017 as a five-day PCAP capture from a controlled, enterprise-like network with benign user activity and scripted attacks.
Although most subsequent studies rely on the CICFlowMeter CSV export, the raw PCAPs reveal a fundamental labeling issue: the public attack description lists external attacker addresses, whereas the captured subnet traffic is post-NAT.
A literal implementation of the public address rules can therefore miss attack flows or assign them to the wrong class.

Liu et al.~\cite{liu2022errorprevalence} audited the dataset and identified empty-payload flows, startup and teardown artifacts, benign background traffic inside attack windows, and incorrect attack-family assignments.
They also reported severe web-attack corruption when destination port 80 is used as a filter.
The lycos-ids2017 project~\cite{lycos2017} provides a corrected reference implementation that uses the internal attacker address and relaxes web-attack filtering to TCP protocol matching.
Our pipeline follows these corrections, adds a Heartbleed source-port constraint to avoid labeling benign TLS flows in the same time window as the attack, and resolves the Thursday Infiltration window by merging labels from a curated exact-match label CSV---assigning attack labels only to flows that match the correct 5-tuple and timestamp bounds rather than blanket-excluding the afternoon period. 

\subsection{Machine Learning Results on CIC-IDS2017}

The broader tabular-learning literature motivates the model families considered by numerous prior studies on variants of CIC-IDS2017 dataset. these studies have evaluated random forests, gradient-boosted trees, support-vector models, neural networks, and ensembles. Gradient-boosted trees are strong baselines for heterogeneous numerical tables~\cite{grinsztajn2022tree}, while neural tabular models such as TabNet~\cite{arik2021tabnet} offer learned feature selection but can be sensitive to imbalance and limited class support. Meta-learning model approach has also been studied recently~\cite{martinez2024redefining}.

Tree ensembles, especially Random Forest and XGBoost, often report accuracy or weighted-F1 above 99\%~\cite{thockchom2023nids}; boosted ensembles have also been reported with high multiclass scores~\cite{okey2022boostedenml}.
A more complex ensemble method based on cognitive diversity of different models was studied in~\cite{owusu2025generalizable}.

The commonly reported weighted-F1 and accuracy are useful diagnostics, but they can remain high even when rare attacks are poorly detected.
Such numbers are difficult to interpret when they come from uncorrected labels, binary-only tasks, random splits without clear class controls, or metrics dominated by benign traffic. A corrected, common evaluation is therefore needed before drawing conclusions about which model families are genuinely effective on CIC-IDS2017. We therefore use macro-F1 as the primary metric: each class contributes equally, so failures on the rare classes including Bot, Web Attack, Heartbleed, or SQL Injection remain visible even when weighted-F1 is high. 

\subsection{Multiclass Evaluation Under Imbalance}

Fine-grained intrusion detection is a multiclass problem with class frequencies that differ by several orders of magnitude.
Linear models commonly use one-vs-rest or one-vs-one reductions~\cite{rifkin2004multiclass}, whereas tree ensembles and neural networks usually optimize multiclass objectives directly.
Regardless of the learner, severe imbalance creates the familiar accuracy paradox: a classifier can score well by emphasizing the majority BENIGN class while missing rare but operationally important attacks~\cite{japkowicz2011imbalanced}.

Common remedies include cost-sensitive learning and oversampling.
SMOTE and related methods synthesize minority-class examples~\cite{chawla2002smote}, while class weighting changes the loss contribution of rare classes.
As He and Garcia~\cite{he2009imbalanced} note, no imbalance strategy is uniformly best. 

We thereafter trained our model for multi-tier classification on binary, attack family, and individual attack classification. The tiered detection granularity is of practical importance in minimizing the security risk and operational cost. 

\subsection{Feature Analysis and Interpretability}
It is common that features collected in a given dataset may have varying dependency and importance despite that all ML models assume feature independency~\cite{yakubu2026automated}. Efficient feature analysis may help reduce overfitting due to redundant features and increase training efficiency from removing the non-relevant features. 

In addition to detection performance, interpretability of the detection models is equally important as attack detection. It makes results and output trusted and comprehensible for both human operators and downstream attack response and mitigation tasks. The classification results can normally be explained through features and their relative importance analysis~\cite{wei2023xnids}.

\section{Dataset and Labeling Methodology}
\label{sec:dataset}

\subsection{CIC-IDS2017 Overview}

The dataset covers five weekdays in July 2017.
Monday contains only benign background traffic; Tuesday through Friday
introduce fourteen attack scenarios executed from a Kali Linux host into an emulated enterprise subnet.
The attack families span denial-of-service (DoS Hulk, DoS GoldenEye,
DoS slowloris, DoS Slowhttptest), volumetric DDoS, port scanning,
brute-force credential attacks (FTP-Patator, SSH-Patator), web
application attacks (Brute Force, XSS, SQL Injection), botnet activity,
network infiltration, and the Heartbleed TLS exploit.
Most attacks are launched from an external Kali Linux host (appearing
internally as \texttt{172.16.0.1} after NAT); the Infiltration scenario is
an exception, where an already-compromised internal host
(\texttt{192.168.10.8}) initiates outbound exfiltration.
After labeling and class-stratified splitting, the test partition contains
372,191 flows are distributed across fifteen classes as shown in Table~\ref{tab:class_dist}. The table also shows the mapping to the three-level class hierarchy: binary (BENIGN vs. ATTACK) and family (coarse group).

\begin{table}[h]
  \caption{Test-split class distribution (15\% of 2,481,273 total flows), with binary-family-class hierarchy.}
  \label{tab:class_dist}
  \centering
  \small
  \setlength{\tabcolsep}{4pt}
  \begin{tabular}{@{}lllcc@{}}
    \toprule
    Binary & Family & Class & Count & Fraction \\
    \midrule
    BENIGN & BENIGN & BENIGN              & 294,480 & 79.12\% \\
    ATTACK & DoS    & DoS Hulk            &  33,436 &  8.98\% \\
    ATTACK & PortScan & PortScan          &  23,892 &  6.42\% \\
    ATTACK & DDoS   & DDoS                &  14,194 &  3.81\% \\
    ATTACK & DoS    & DoS GoldenEye       &   1,434 &  0.39\% \\
    ATTACK & DoS    & DoS slowloris       &   1,329 &  0.36\% \\
    ATTACK & DoS    & DoS Slowhttptest    &   1,363 &  0.37\% \\
    ATTACK & BruteForce & FTP-Patator      &     963 &  0.26\% \\
    ATTACK & BruteForce & SSH-Patator      &     451 &  0.12\% \\
    ATTACK & Bot    & Bot                 &     331 &  0.09\% \\
    ATTACK & WebAttack & Web Attack BF     &     204 &  0.05\% \\
    ATTACK & WebAttack & Web Attack XSS    &      99 &  0.03\% \\
    ATTACK & Infiltration & Infiltration    &      10 & $<$0.01\% \\
    ATTACK & WebAttack & Web Attack SQLi   &       4 & $<$0.01\% \\
    ATTACK & Heartbleed & Heartbleed       &       1 & $<$0.01\% \\
    \bottomrule
  \end{tabular}
\end{table}

The imbalance ratio between the majority class (BENIGN) and the rarest
attack classes exceeds $10^5$, placing this benchmark firmly in the
extreme-imbalance regime.

\subsection{Flow Labeling}

Labeling starts from the raw PCAP files for each capture day.
We assemble bidirectional flows using the 5-tuple (source IP, destination IP, source port, destination port, protocol), a 30-second idle timeout, and a 120-second active timeout.
Each flow is matched to corrected label records using timestamp-aware 5-tuple exact matching (source/destination IP, source/destination port, protocol), with schedule-derived windows used as metadata in the label records.
After the corrections below, each flow receives a label from the 15-class taxonomy; the resulting table is then split into training, validation, and test sets.
The labeling code and configuration are included in the project repository.

We correct five issues that affect common CIC-IDS2017 labeling workflows:

\textbf{NAT translation.}
The UNB documentation lists the attacker as \texttt{205.174.165.73}, but this address undergoes source NAT at the gateway firewall before entering
the captured subnet.
Attacker-initiated flows (DoS, DDoS, brute-force, port scan, web attacks), therefore, carry the firewall's internal interface address
\texttt{172.16.0.1} as their source IP in the PCAPs.
Labels that filter on the public address miss these flows entirely; our
corrected label CSV uses the post-NAT address for all such attack classes.
Note that Infiltration is an exception: it represents an internal
compromised host (\texttt{192.168.10.8}) exfiltrating data outward, so
its flow direction and IP assignment differ from the other attack classes.

\textbf{Web attack port filter.}
Several published implementations, including an earlier version of the
Lycos reference, filter web-attack flows by destination port 80.
Liu et al.~\cite{liu2022errorprevalence} measured 95.16\% corruption in
Web Brute Force and 94.48\% in Web XSS under this condition because many
legitimate attack flows arrive on other ports during page navigation, form
submission, and redirect handling.
The corrected implementation filters by TCP protocol (\texttt{ip\_prot=6})
only, matching the final Lycos reference behavior.

\textbf{FTP-Patator and SSH-Patator port filters.}
An early version of our labeling script required destination ports 21 and
22, respectively, for the patator attacks.
The Lycos reference uses only timestamp and IP address matching, adding
port filters miss flows where the attacker probes non-standard ports as
part of the credential-stuffing sweep.
Both port requirements were removed.

\textbf{Heartbleed source port.}
The Heartbleed traffic in CIC-IDS2017 is generated from a fixed ephemeral
source port (45022) toward the victim's TLS listener on port 444.
Without the source-port constraint, benign traffic toward port 444 is
captured inside the attack time window and mislabeled.
Adding the source-port filter to the matching predicate reduces
false positives on this single-sample class.

\textbf{Infiltration Thursday handling.}
The Thursday scenario includes web attacks in the morning and an
infiltration episode later in the day.
Window-only heuristics can mix benign and attack traffic in this period,
which is why prior audits report contamination risk~\cite{liu2022errorprevalence}.
In our current pipeline, exact flow-key matching
(timestamp + source/destination IP + source/destination port + protocol) is used, 
then mapped to the canonical taxonomy.
Under this exact-match process, infiltration attack flows are cleanly
retained in the exported flow CSV; we do not apply a blanket exclusion of
Thursday afternoon traffic.

\subsection{Feature Set}

Each labeled flow is represented by 71 numerical statistics computed directly from the PCAP-derived bidirectional flow record.
The features summarize packet and byte counts, flow duration and inter-arrival times, packet lengths, TCP flags, throughput, header and protocol metadata, bulk-transfer behavior, subflow statistics, and active/idle timing.
These nine groups capture the volume, timing, directionality, and connection-state patterns needed to distinguish floods, scans, credential attacks, and web sessions.

\begin{itemize}
  \item \textbf{Packet Counts and Byte Volumes (6 features):} Total packets and bytes sent and received in each direction, and their sums. These features capture the overall size and directionality of the flow, distinguishing, for example, large file transfers from short web requests.
  \item \textbf{Flow Duration and Inter-Arrival Time Statistics (13 features):} Flow duration plus mean, standard deviation, min, and max of packet inter-arrival times in the forward, backward, and combined directions. These features are sensitive to bursty or periodic traffic patterns, which are characteristic of certain attacks (e.g., DoS floods) and benign behaviors (e.g., streaming).
  \item \textbf{Packet-Length Statistics (12 features):} Statistical summaries (mean, std, min, max, quantiles) of packet sizes in each direction. These help differentiate between protocols and application types and detect anomalies such as fixed-size attack payloads.
  \item \textbf{TCP Flag Counts (10 features):} Counts of SYN, ACK, FIN, RST, PSH, URG, ECE, CWR, and flag combinations. These features are critical for identifying scanning, connection attempts, and protocol misuse.
  \item \textbf{Throughput Rates (5 features):} Bytes and packets per second, computed over the flow duration and in each direction. These features highlight high-rate attacks and distinguish them from normal user activity.
  \item \textbf{Header and Protocol Metadata (5 features):} Protocol type (TCP/UDP/ICMP), source and destination ports, and forward/backward header lengths. These provide context for interpreting the other statistics and enable protocol-aware discrimination (e.g., separating TCP-based attacks from UDP floods).
  \item \textbf{Bulk-Transfer Statistics (8 features):} Metrics such as the number and size of bulk data transfers, bulk duration, and bulk packet counts. These are designed to capture behaviors such as file downloads, exfiltration, and large uploads.
  \item \textbf{Subflow Statistics (4 features):} Packet and byte counts for the forward and backward subflows within the flow. These capture finer-grained directionality information that complements the top-level byte and packet counts.
  \item \textbf{Active/Idle Time Statistics (8 features):} Mean, standard deviation, min, and max of active and idle period durations within the flow, measuring how traffic is distributed over time. These features help distinguish between interactive sessions and automated or scripted activity.
\end{itemize}

\section{Data Processing and Modeling Pipeline}
\label{sec:pipeline}

The full workflow from raw PCAP to model evaluation is implemented as a reproducible pipeline with the following steps:

\begin{enumerate}
  \item \textbf{Flow Assembly and Labeling:} Raw PCAPs are parsed to extract bidirectional flows using a 5-tuple key and timeouts. Each flow is labeled using the corrected exact-match label CSV described in Section~\ref{sec:dataset}, producing a single labeled table.
  \item \textbf{Feature Extraction:}  Feature extraction is performed independently for each completed flow, with no context carried across flows or future windows.
Metadata needed for labeling and auditing is removed before modeling: \texttt{flow\_id}, \texttt{src\_ip}, \texttt{dst\_ip}, \texttt{flow\_start\_time}, \texttt{scenario\_id}, and \texttt{attack\_class}.
This prevents leakage from identifiers, timestamps, scenario membership, labels, or split membership.
  \item \textbf{Preprocessing:} Missing values are imputed with per-feature medians computed on the training split. Features are standardized (zero mean, unit variance) using a scaler fit only to the training split. Categorical metadata columns are excluded. All preprocessing objects are persisted for reproducibility.
  \item \textbf{Data Splitting:} The labeled and preprocessed flows are partitioned into train (70\%), validation (15\%), and test (15\%) sets using a seeded stratified shuffle on the coarse attack family. This ensures all classes are represented in each split and prevents leakage.
  \item \textbf{Model Training and Evaluation:} Eleven classifiers are trained on the training split, with hyperparameters set curated from prior literature. The model suite covers:
    \begin{itemize}
      \item \textbf{Linear:} Logistic Regression (LR, $L_2$ regularization, balanced class weights, L-BFGS solver) serves as a lower-bound reference, valued for speed and interpretability but limited by its inability to capture non-linear boundaries between attack families.
      \item \textbf{Shallow Bagging Ensembles:} Random Forest (RF, 300 trees, balanced class weights) and Extra Trees (ET, 300 trees) use bootstrap resampling and random feature selection to reduce variance and capture feature interactions. AdaBoost (200 stumps) sequentially reweights errors but is limited by its weak, axis-aligned base learners.
      \item \textbf{Gradient-Boosted Trees:} XGBoost (XGB, 500 trees, depth 6, learning rate 0.05), LightGBM (LGBM, 300 trees, 31 leaves), and CatBoost (500 iterations, MultiClass loss, balanced class weights) all use boosting to learn non-linear, high-dimensional boundaries and handle class imbalance. CatBoost's aggressive class balancing can over-correct for rare classes.
      \item \textbf{Neural/Attention-Based:} MLP (three-layer, 256--256--128 units, ReLU) is a universal function approximator but requires many samples per class. TabNet uses sequential attention to select features at each decision step, mimicking tree splits.
      \item \textbf{Meta-Learners:} Stacking combines XGBoost, MLP, and Logistic Regression as base estimators, with a Logistic Regression meta-classifier trained on their out-of-fold predictions. This maximizes diversity and tests whether a learned combination outperforms simple averaging. Soft voting averages the class probability vectors from Random Forest, XGBoost, and LightGBM to test whether combining strong but similar models is more effective. In our results, the soft-voting ensemble outperforms stacking, suggesting that averaging the outputs of the strongest tree-based models is most effective for this dataset.
    \end{itemize}
    The models also differ in how much intrinsic imbalance handling they provide.
Logistic Regression, Random Forest, and several boosting implementations can use class weights or sample weights; tree ensembles partially reduce variance through averaging but do not remove minority-class scarcity; gradient-boosted trees can focus subsequent trees on hard examples, although this can over-emphasize rare classes if calibration is poor; neural models have no inherent protection against imbalance unless the loss or sampler is modified. Models are evaluated on the test split using macro-averaged and weighted F1 at three granularity levels (binary, coarse, fine). Per-class metrics, confusion matrices, and bar charts are generated for detailed analysis. All code, configuration, and artifacts are available for full reproducibility.
\end{enumerate}

This unified pipeline ensures that every result is traceable from raw data to final metric, and that all preprocessing, splitting, and evaluation steps are leakage-safe and reproducible. The pipeline design supports extension to new datasets and model families with minimal changes.

\subsection{Evaluation Metrics}
\label{sec:metrics-guide}

Evaluation is reported at three nested tiers: binary (BENIGN versus ATTACK), coarse family (nine classes), and fine class (fifteen classes).
For a class $c$, F1 is the harmonic mean of precision and recall:
\begin{equation*}
  \text{F1}_c = \frac{2 \cdot \text{Precision}_c \cdot \text{Recall}_c}{\text{Precision}_c + \text{Recall}_c}.
\end{equation*}
Macro-F1 averages class-level F1 scores uniformly,
\begin{equation*}
  \text{Macro-F1} = \frac{1}{K} \sum_{c=1}^K \text{F1}_c,
\end{equation*}
where $K$ is the number of classes.
Weighted-F1 instead weights each class by test support,
\begin{equation*}
  \text{Weighted-F1} = \frac{1}{N} \sum_{c=1}^K n_c \cdot \text{F1}_c,
\end{equation*}
where $n_c$ is the number of test samples in class $c$ and $N$ is the test-set size.
Because BENIGN flows dominate CIC-IDS2017, macro-F1 is the primary metric, and weighted-F1 is reported only to show how much majority-class performance can inflate aggregate scores.
Confusion matrices and per-class F1 plots are used to identify which attack classes drive the aggregate results.

\subsection{Feature Importance, Feature Selection, and Explainability}
\label{sec:feature_importance}
Feature importance analysis is key to the explainability of developed ML models, and is also used in model and feature selection. There are two major methods: the internal method in Scikit-learn models, which measures internal model metrics to yield global importance~\cite{tavares2025variability}, and the SHAP method, which yields local importance for individual predictions, which can then be averaged to create global rankings~\cite{yakubu2026automated}. SHAP is mathematically guaranteed to be consistent across different model types and shows both the magnitude and the direction (positive/negative impact) of each feature. For this study, we use the native global feature importance score and ranking and conduct an ablation study by removing the least important features.

\section{Results}
\label{sec:results}

\subsection{Overall Model Comparison}

Table~\ref{tab:model_comparison} reports the main test-split results.
The central finding is stable across all three tiers: tree ensembles dominate the tabular CIC-IDS2017 task, and the soft-voting ensemble is the strongest overall model.
Voting achieves 0.955 macro-F1 at the fine fifteen-class tier, followed closely by XGBoost (0.952), LightGBM (0.948), and Random Forest (0.932).
Below this group, performance drops substantially: Extra Trees reaches 0.895, Stacking reaches 0.817, TabNet reaches 0.720, CatBoost reaches 0.666, and the remaining baselines fall to 0.560 or lower.

\begin{table}[h]
  \caption{Test-split performance across three evaluation tiers.
           Macro-F1 (M-F1) and weighted-F1 (W-F1) are shown.
           Bold indicates the best value in each column.}
  \label{tab:model_comparison}
  \centering
  \small
  \setlength{\tabcolsep}{3pt}
  \begin{tabular}{@{}lcccccc@{}}
    \toprule
    \multirow{2}{*}{Model}
      & \multicolumn{2}{c}{Binary}
      & \multicolumn{2}{c}{Coarse (9)}
      & \multicolumn{2}{c}{Fine (15)} \\
    \cmidrule(lr){2-3}\cmidrule(lr){4-5}\cmidrule(lr){6-7}
      & M-F1 & W-F1
      & M-F1 & W-F1
      & M-F1 & W-F1 \\
    \midrule
    Voting       & \textbf{.999} & \textbf{.999} & \textbf{.980} & \textbf{.999} & \textbf{.955} & \textbf{.999} \\
    XGBoost      & .996          & .997          & .976          & .997          & .952          & .997          \\
    LightGBM     & .997          & .998          & .979          & .998          & .948          & .998          \\
    Random Forest& .996          & .998          & .967          & .998          & .932          & .998          \\
    Extra Trees  & .996          & .997          & .941          & .997          & .895          & .997          \\
    Stacking     & .993          & .990          & .953          & .991          & .817          & .990          \\
    TabNet       & .982          & .988          & .928          & .988          & .720          & .988          \\
    CatBoost     & .978          & .974          & .909          & .974          & .666          & .974          \\
    MLP          & .960          & .966          & .850          & .966          & .560          & .966          \\
    Log.\ Reg.   & .944          & .921          & .812          & .921          & .506          & .921          \\
    AdaBoost     & .912          & .852          & .781          & .852          & .439          & .852          \\
    \bottomrule
  \end{tabular}
\end{table}

The table also shows why macro-F1 is necessary.
Weighted-F1 remains high for most models because BENIGN accounts for 79\% of the test set and is easy to separate.
For example, the voting ensemble scores 0.999 weighted-F1 but 0.955 macro-F1 at the fine tier; the difference is the minority-class penalty that weighted averages hide.
The fine-tier degradation is small for the best boosted models but large for weaker learners, indicating that binary detection is easier than reliably achieving fine-grained attribution.
This pattern is consistent with the imbalance properties of the model families: methods that can exploit class weighting, hard-example fitting, and ensemble averaging retain more minority-class signal, while linear, shallow-boosting, and neural baselines degrade when rare classes have limited support.

\subsection{Per-Class Behavior of the Top Models}

Table~\ref{tab:perclass_top3} gives the most useful numerical detail: per-class F1 for Voting, XGBoost, and LightGBM.
Most classes are effectively solved by these models, including BENIGN, DDoS, FTP-Patator, SSH-Patator, and the major DoS variants.
The unresolved errors are concentrated in four low-support classes: Bot, Web Attack Brute Force, Web Attack XSS, and Web Attack SQL Injection.

\begin{table}[h]
  \caption{Per-class F1 for the three best models. ``BF'' = Brute Force,
           ``XSS'' = Cross-Site Scripting, ``SQLi'' = SQL Injection.}
  \label{tab:perclass_top3}
  \centering
  \small
  \setlength{\tabcolsep}{3pt}
  \begin{tabular}{@{}lccc@{}}
    \toprule
    Class & Voting & XGBoost & LightGBM \\
    \midrule
    BENIGN              & .999 & .998 & .999 \\
    DoS Hulk            & 1.000 & .999 & 1.000 \\
    PortScan            & .995 & .981 & .989 \\
    DDoS                & 1.000 & 1.000 & 1.000 \\
    DoS GoldenEye       & .998 & .996 & .997 \\
    FTP-Patator         & 1.000 & 1.000 & 1.000 \\
    SSH-Patator         & .999 & .999 & 1.000 \\
    DoS slowloris       & .998 & .998 & .998 \\
    DoS Slowhttptest    & .995 & .993 & .995 \\
    Bot                 & .891 & .875 & .885 \\
    Web Attack BF       & .872 & .862 & .875 \\
    Web Attack XSS      & .772 & .774 & .785 \\
    Web Attack SQLi     & .857 & .857 & .750 \\
    Infiltration        & .947 & .947 & .947 \\
    Heartbleed          & 1.000 & 1.000 & 1.000 \\
    \midrule
    \textbf{Macro-F1}   & \textbf{.955} & \textbf{.952} & \textbf{.948} \\
    \bottomrule
  \end{tabular}
\end{table}

Bot and web attacks remain the difficult cases.
Because host identifiers, timestamps, scenario identifiers, and labels are excluded, the models must rely only on flow-level statistical behavior.
Bot traffic overlaps benign HTTP browsing in packet length, timing, and throughput, yielding high recall but lower precision.
Web Attack, Brute Force, and XSS are also close to normal web sessions in the features they retain; only subtle differences in rate, size, flag, and response patterns remain.
SQL Injection and Heartbleed have only four and one test samples, respectively, so their scores are reported for completeness rather than as stable estimates.

\subsection{Confusion Matrix and Per-Class Figures}

\begin{figure}[h]
  \centering
  \includegraphics[width=\columnwidth]{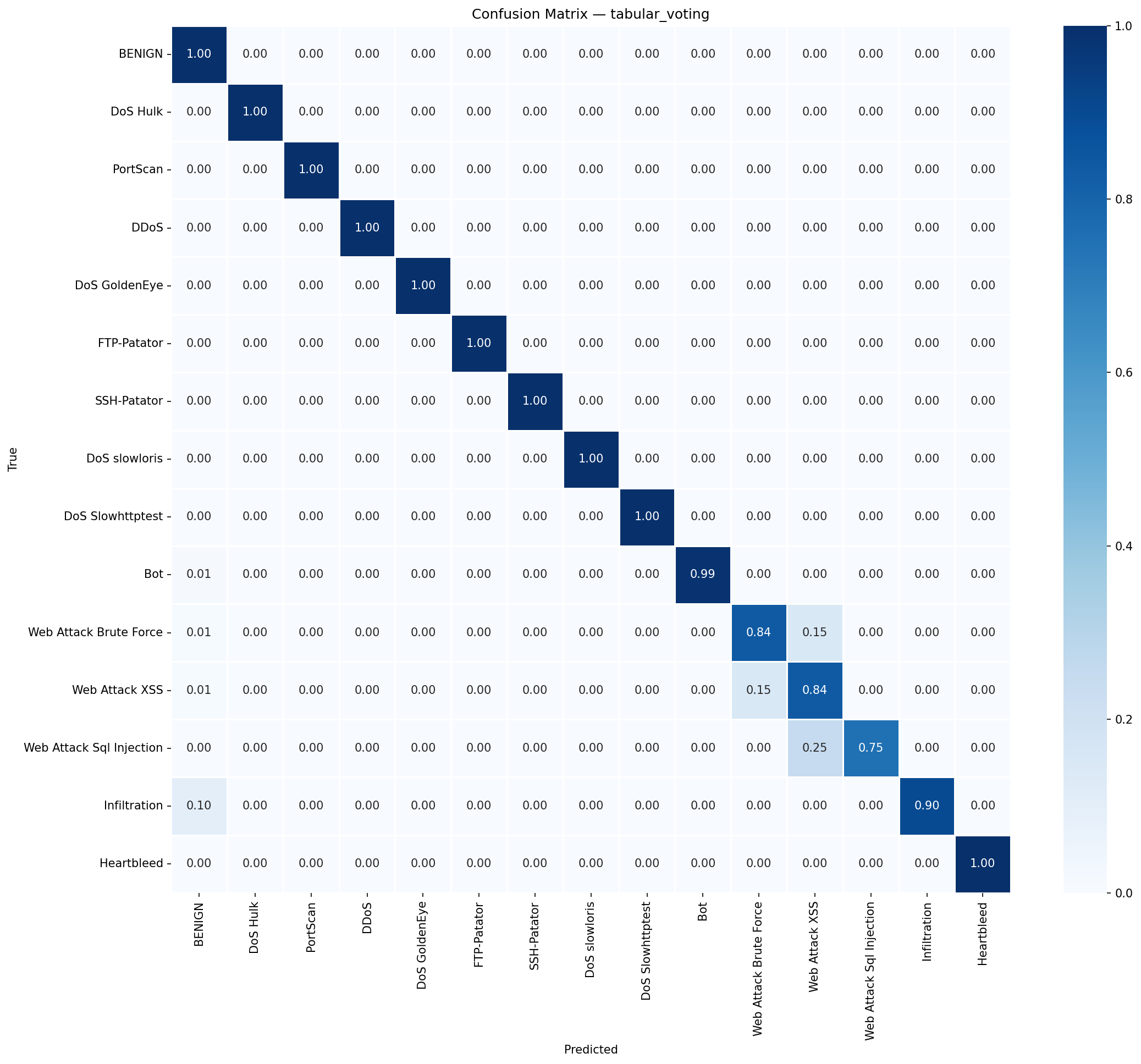}
  \caption{Confusion matrix for the soft-voting ensemble on the 15-class
           test split (log-scale color mapping).}
  \label{fig:cm_voting}
\end{figure}

Fig.~\ref{fig:cm_voting} summarizes the fine-tier errors for the best model.
The largest visible off-diagonal block is PortScan predicted as BENIGN, while Bot and Web Attack errors also mostly fall into BENIGN.
The DoS subtypes are cleanly separated, suggesting that the extracted timing, packet length, and flag count features are sufficient for these attacks.

\begin{figure}[h]
  \centering
  \includegraphics[width=\columnwidth]{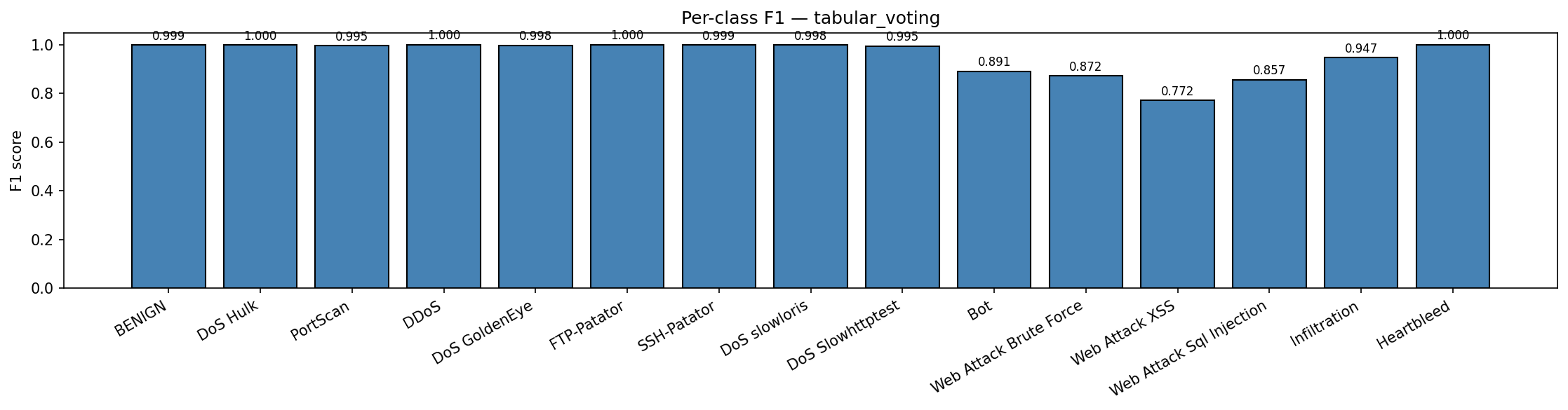}
  \caption{Per-class F1 for the soft-voting ensemble.}
  \label{fig:f1_voting}
\end{figure}

\begin{figure}[h]
  \centering
  \includegraphics[width=\columnwidth]{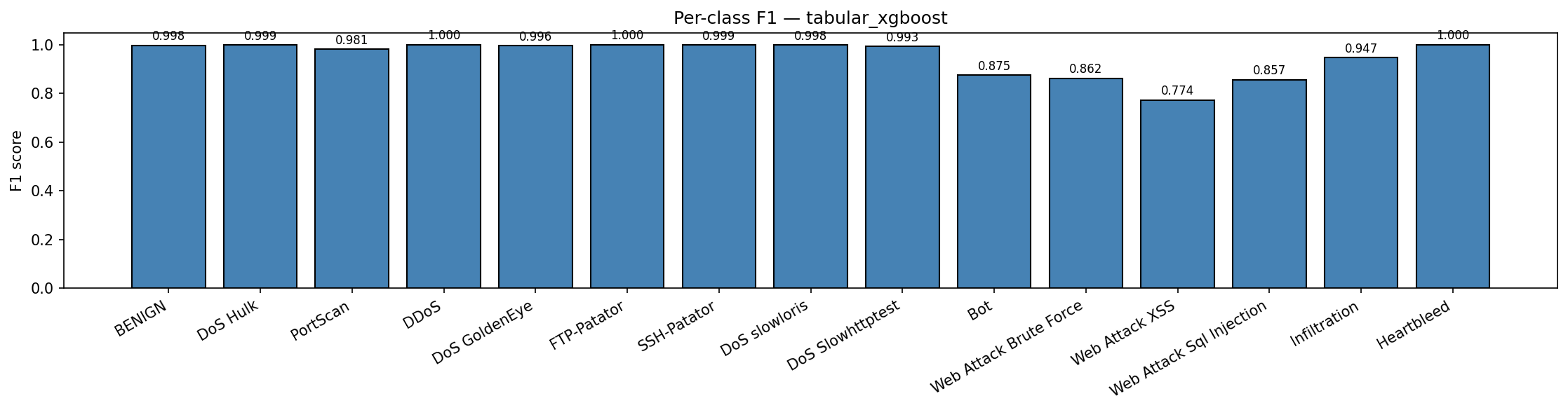}
  \caption{Per-class F1 for XGBoost.}
  \label{fig:f1_xgb}
\end{figure}

\begin{figure}[h]
  \centering
  \includegraphics[width=\columnwidth]{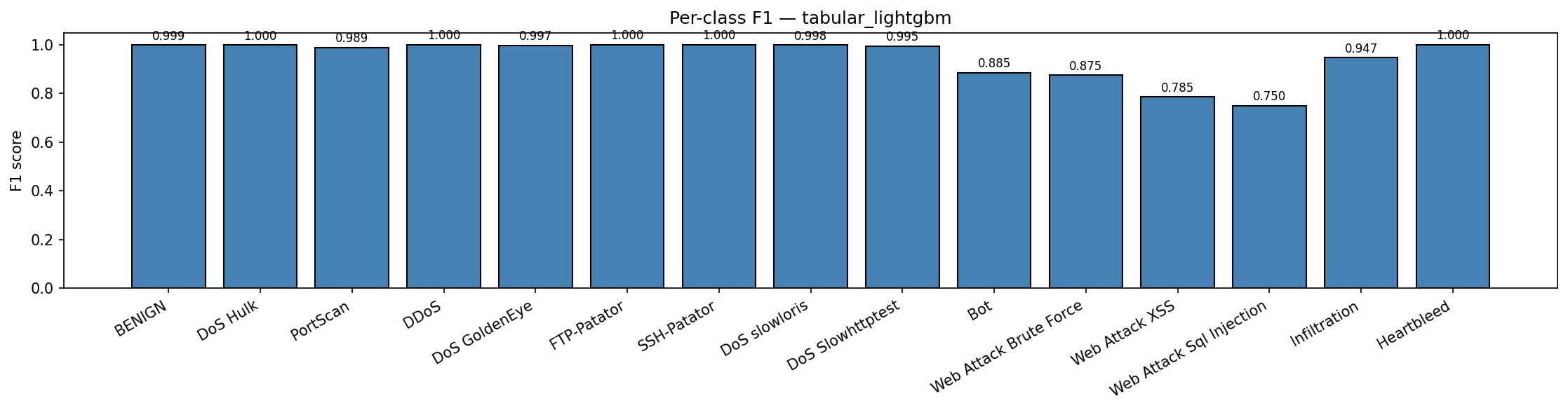}
  \caption{Per-class F1 for LightGBM.}
  \label{fig:f1_lgbm}
\end{figure}

Fig.~\ref{fig:f1_voting}, Fig.~\ref{fig:f1_xgb}, and Fig.~\ref{fig:f1_lgbm} show that the three leading models share the same performance profile.
The consistency across these models indicates that the remaining errors are not idiosyncratic to one classifier; they reflect the difficulty of separating rare bot and web-attack flows from benign traffic using only tabular flow features.

\subsection{Feature Importance Analysis and Feature Selection}

Feature importance analysis of the soft voting model, the highest-performing model, reveals that several features have little or no importance. The 3 least-important features have zero importance score: $cwe\_flag\_count$ (Congestion Window Reduced Flag Count)
$ece\_flag\_count$ (Explicit Congestion Notification-Echo Flag Count), and 
$bwd\_urg\_flag_count$ (Backward Urgent Flag Count). All these flags measure uncommon TCP control flags
Rarely appear in typical network traffic
Provide little information for distinguishing attacks from benign traffic. On the other hand, the three most important features are:
$bwd\_pkt\_len\_max$ (Maximum Backward Packet Length) - large/unusual responses indicating abnormal traffic behavior; $bwd\_pkt\_len\_min$ (Minimum Backward Packet Length) - very small response packets can also reveal characteristic communication patterns used by certain attacks; $protocol$ - TCP, UDP, or another protocol identifiers help distinguish different families of network attacks.

We further conducted an ablation study by removing the least important features and re-running the whole pipeline. The results in Table~\ref{fig:feature_selection} show that removing the bottom three features has a negligible effect on the classification performance. It is interesting to note that removing the bottom four or nine features negatively affects Macro-F1 performance, albeit by a similar margin. Fewer features apparently save training time.  

\begin{figure}[h]
\centering

\begin{subfigure}{0.95\linewidth}
    \centering
    \small

    \begin{tabular}{lcc}
        \toprule
        \multicolumn{3}{c}{\textbf{Voting Without Bottom 3 Features (71--69)}} \\
        \midrule
        \textbf{Metric} & \textbf{Value} & $\boldsymbol{\Delta}$ \\
        \midrule
        Accuracy      & 0.998297 & -0.000026 \\
        Macro-F1      & 0.946564 & -0.000419 \\
        Weighted-F1   & 0.998315 & -0.000026 \\
        Training Time & 0:43:46  & -0:02:03 \\
        \bottomrule
    \end{tabular}

    \caption{Removing bottom three features}
    \label{fig:feature_3}
\end{subfigure}

\vspace{0.6em}

\begin{subfigure}{0.95\linewidth}
    \centering
    \small

    \begin{tabular}{lcc}
        \toprule
        \multicolumn{3}{c}{\textbf{Voting Without Bottom 4 Features (71--68)}} \\
        \midrule
        \textbf{Metric} & \textbf{Value} & $\boldsymbol{\Delta}$ \\
        \midrule
        Accuracy      & 0.998388 & +0.000065 \\
        Macro-F1      & 0.925532 & -0.021451 \\
        Weighted-F1   & 0.998409 & +0.000068 \\
        Training Time & 0:41:50  & -0:03:59 \\
        \bottomrule
    \end{tabular}

    \caption{Removing bottom four features}
    \label{fig:feature_4}
\end{subfigure}

\vspace{0.6em}

\begin{subfigure}{0.95\linewidth}
    \centering
    \small

    \begin{tabular}{lcc}
        \toprule
        \multicolumn{3}{c}{\textbf{Voting Without Bottom 9 Features (71--63)}} \\
        \midrule
        \textbf{Metric} & \textbf{Value} & $\boldsymbol{\Delta}$ \\
        \midrule
        Accuracy      & 0.998302 & -0.000021 \\
        Macro-F1      & 0.945055 & -0.001928 \\
        Weighted-F1   & 0.998320 & -0.000021 \\
        Training Time & 0:41:32  & -0:04:17 \\
        \bottomrule
    \end{tabular}

    \caption{Removing bottom nine features}
    \label{fig:feature_9}
\end{subfigure}

\caption{Feature Selection}
\label{fig:feature_selection}
\end{figure}

\subsection{Implications of Multi-Tier Evaluation}

The three-tier design separates the solved parts of CIC-IDS2017 from the remaining hard cases.
At the binary tier, nearly all models except AdaBoost exceed 0.96 macro-F1.
At the coarse tier, the tree ensembles remain strong, with most of the degradation occurring in the Bot and WebAttack families.
At the fine tier, only the strongest boosted or ensemble models maintain macro-F1 above 0.93.
Binary performance should therefore not be used as evidence of fine-grained attribution quality.
For the corrected tabular CIC-IDS2017, the benchmark now serves as a controlled reference for tiered class and feature analysis that can be used for  also cross-dataset transfer.

\section{Discussion}
\label{sec:discussion}

\subsection{What the Benchmark Establishes}

The corrected pipeline removes several sources of artificial performance inflation, making the model comparison more reliable.
With this protocol, the ranking is clear: XGBoost, LightGBM, Random Forest, and their soft-voting ensemble are the strongest tabular classifiers, while linear, shallow-boosting, and neural-tabular baselines degrade on the fifteen-class task.
Table~\ref{tab:model_comparison}, Table~\ref{tab:perclass_top3}, and Fig.~\ref{fig:cm_voting} show that performance is effectively saturated for binary detection, coarse-family attribution, and most fine-grained classes.
The remaining fine-tier errors are concentrated in classes with very small test support or flow-feature distributions that are close to those of benign web traffic.
For single-dataset CIC-IDS2017 studies, this leaves little value in simply substituting another tabular classifier; meaningful progress requires changes to the data, feature representation, rare-class treatment, or transfer setting.

\subsection{Limitations}

CIC-IDS2017 is still a controlled, five-day benchmark rather than a deployment trace.
Its traffic generator, topology, attack tools, and attacker address patterns are dataset-specific, so corrected labels alone do not establish operational transfer.
The smallest classes also limit statistical confidence: Heartbleed has one test sample, SQL Injection has four, and Infiltration has ten.
These classes are useful for verifying that a pipeline preserves rare labels, but not for estimating real-world detector performance with precision.

\subsection{Future Work and Deployment}

The main technical opportunity is rare-class discrimination without leakage.
Because the leading tree-based models already combine class weighting, hard-example fitting, and ensemble averaging, synthetic oversampling is unlikely to change the headline conclusion on CIC-IDS2017.
Methods such as SMOTE may still be useful as rare-class ablations, but for classes with only one to four test samples, they cannot substitute for additional labeled data.
Class-specific calibration and cost-sensitive threshold tuning could further improve Bot and Web Attack precision--recall trade-offs without changing the training distribution.
Application-layer features such as HTTP request structure, TLS handshake behavior, or DNS entropy may also be needed to separate minority web and bot traffic from benign flows.
Cross-dataset evaluation on newer testbed and enterprise traces is necessary to test whether the learned flow-level patterns transfer beyond CIC-IDS2017~\cite{sharafaldin2018toward,mohammadian2024poisoning}.

For deployment, the three-tier hierarchy naturally supports a staged response.
A fast binary detector can act as a gate, a coarse-family model can route alerts to an appropriate response workflow, and fine-class attribution can enrich tickets when confidence is high.
In practice, the bottleneck is less likely to be tree-model inference than upstream flow assembly and feature extraction at line rate.
Operational use would therefore require a streaming feature pipeline, drift monitoring, periodic retraining on analyst-confirmed alerts, and promotion only when minority-class performance is preserved.

For future networks, the main implication is that AI-based NIDS benchmarks must be leakage-safe and reproducible before they are used to justify automated security decisions in 5G/6G, edge, IoT, or SDN/NFV environments.
Future-network traffic will differ from CIC-IDS2017, but the evaluation failures exposed here---label ambiguity, testbed-specific metadata, majority-class metrics, and rare-class instability---are general.
A comprehensive benchmark such as this one is therefore useful as a  reference workflow for trustworthy IDS evaluation in emerging network settings.

\section{Conclusion}
\label{sec:conclusion}

This paper presented a corrected benchmark for tabular, multi-class, multi-tier intrusion detection on the CIC-IDS2017 dataset.
Across eleven classifiers, a soft-voting ensemble of Random Forest, XGBoost, and LightGBM gives the best overall result, reaching 0.955 macro-F1 on the fifteen-class task.
The strongest individual models are close behind, and the leading tree-based models achieve nearly perfect binary and coarse-family detection.
At fine resolution, most remaining errors involve rare Bot and Web Attack classes, where support is limited and flow-only features are intrinsically ambiguous.
These results indicate that purely single-dataset tabular studies on CIC-IDS2017 have reached a natural endpoint: the dataset remains valuable as a reproducible benchmark and pipeline reference, but future progress should come from rare-class support, richer features, and cross-dataset validation rather than another isolated model comparison on the same data.

\balance

\section*{Acknowledgment}

The authors thank the Canadian Institute for Cybersecurity (CIC) at the
University of New Brunswick for releasing the CIC-IDS2017 dataset and
associated traffic characterization documentation. The authors also acknowledge 
the LYCOS team at the University of Le Mans for publishing their reference labeling implementation.

The authors used the NSF FABRIC testbed for data hosting, software development, and evaluation,
GitHub Copilot in Visual Studio Code for code development assistance, and OpenAI Prism for manuscript polishing.
The authors reviewed and verified all technical content, experimental results, claims, references, and final text, and remain fully responsible for the work.


\begin{thebibliography}{99}

\bibitem{lones2024avoiding}
Lones, Michael A,
"Avoiding common machine learning pitfalls,"
Patterns, vol.\ 5, no.\ 10, 2024, Elsevier.

\bibitem{gorishniy2021revisiting}
Gorishniy, Yury and Rubachev, Ivan and Khrulkov, Valentin and Babenko, Artem,
``Revisiting deep learning models for tabular data,''
in \emph{Proc.\ Advances in neural information processing systems}, 34, 2021, 18932--18943.

\bibitem{sharafaldin2018cicids}
I.~Sharafaldin, A.~H.~Lashkari, and A.~A.~Ghorbani,
``Toward generating a new intrusion detection dataset and intrusion
traffic characterization,''
in \emph{Proc.\ ICISSP}, Funchal, Madeira, Portugal, Jan.\ 2018, pp.\ 108--116.

\bibitem{liu2022errorprevalence}
Z.~Liu, C.~Yin, and Y.~Hu,
``Error prevalence in NIDS datasets: A case study on CIC-IDS-2017 and
CSE-CIC-IDS-2018,''
in \emph{Proc.\ IEEE CNS}, Austin, TX, USA, Oct.\ 2022, pp.\ 1--9.

\bibitem{lycos2017}
University of Le Mans,
``lycos-ids2017 reference labeling implementation,''
\url{https://maupiti-git.univ-lemans.fr/lycos/lycos-ids2017}, 2020.

\bibitem{okey2022boostedenml}
O.~D.~Okey, S.~S.~Maidin, P.~Adasme, R.~L.~Rosa, M.~Saadi,
D.~Melgarejo, and D.~Z.~Rodr\'{i}guez,
``BoostedEnML: Efficient technique for detecting cyberattacks in IoT
systems using boosted ensemble machine learning,''
\emph{Sensors}, vol.\ 22, no.\ 21, p.\ 7409, 2022.

\bibitem{thockchom2023nids}
N.~Thockchom, M.~M.~Singh, and U.~Nandi,
``A novel ensemble learning-based model for network intrusion detection,''
\emph{Complex \& Intelligent Systems}, vol.\ 9, no.\ 5, pp.\ 5693--5714, 2023.

\bibitem{arik2021tabnet}
S.~\"{O}.~Arik and T.~Pfister,
``TabNet: Attentive interpretable tabular learning,''
in \emph{Proc.\ AAAI}, vol.\ 35, no.\ 8, 2021, pp.\ 6679--6687.

\bibitem{japkowicz2011imbalanced}
N.~Japkowicz and M.~Shah,
\emph{Evaluating Learning Algorithms: A Classification Perspective}.
Cambridge, UK: Cambridge Univ.\ Press, 2011.

\bibitem{grinsztajn2022tree}
L.~Grinsztajn, E.~Oyallon, and G.~Varoquaux,
``Why tree-based models still outperform deep learning on tabular data,''
in \emph{Proc.\ NeurIPS}, New Orleans, LA, USA, Dec.\ 2022.

\bibitem{chawla2002smote}
N.~V.~Chawla, K.~W.~Bowyer, L.~O.~Hall, and W.~P.~Kegelmeyer,
``SMOTE: Synthetic minority over-sampling technique,''
\emph{J.\ Artif.\ Intell.\ Res.}, vol.\ 16, pp.\ 321--357, 2002.

\bibitem{he2009imbalanced}
H.~He and E.~A.~Garcia,
``Learning from imbalanced data,''
\emph{IEEE Trans.\ Knowl.\ Data Eng.}, vol.\ 21, no.\ 9, pp.\ 1263--1284,
Sep.\ 2009.

\bibitem{rifkin2004multiclass}
R.~Rifkin and A.~Klautau,
``In defense of one-vs-all classification,''
\emph{J.\ Mach.\ Learn.\ Res.}, vol.\ 5, pp.\ 101--141, Jan.\ 2004.

\bibitem{chen2016xgboost}
T.~Chen and C.~Guestrin,
``XGBoost: A scalable tree boosting system,''
in \emph{Proc.\ ACM KDD}, San Francisco, CA, USA, Aug.\ 2016, pp.\ 785--794.

\bibitem{lundberg2017shap}
S.~M.~Lundberg and S.-I.~Lee,
``A unified approach to interpreting model predictions,''
in \emph{Proc.\ NeurIPS}, Long Beach, CA, USA, Dec.\ 2017, pp.\ 4765--4774.

\bibitem{xin2026repo}
Y. Xin, "network-attack-research: CIC-IDS2017/2018 NIDS research artifacts and reproducibility resources," GitHub repository, 2026. Available: \url{https://github.com/YufengXin/network-attack-research}

\bibitem{tavares2025variability}
  Tavares, Cristina and Nascimento, "Variability-Aware Machine Learning Model Selection: Feature Modeling, Instantiation, and Experimental Case Study,"
  IEEE Access,2025.
\bibitem{yakubu2026automated}
Yakubu, P.B., Santana, L., Rahouti, M., Xin, Y., Chehri, A. and Aledhari, M., "Automated and Explainable Denial of Service Analysis for AI‐Driven Intrusion Detection Systems," IET Information Security, 2026(1), p.7264961.

\bibitem{sharafaldin2018toward}
Sharafaldin, I., Lashkari, A.H. and Ghorbani, A.A., 2018. Toward generating a new intrusion detection dataset and intrusion traffic characterization. ICISSP, 1(2018), pp.108-116.

\bibitem{mohammadian2024poisoning}
Mohammadian, H., Lashkari, A.H. and Ghorbani, A.A., 2024, August. Poisoning and evasion: Deep learning-based nids under adversarial attacks. In 2024 21st Annual international conference on privacy, security and trust (PST) (pp. 1-9). IEEE.

\bibitem{panwar2022intrusion}
Panwar, S.S., Raiwani, Y.P. and Panwar, L.S., 2022, November. An intrusion detection model for cicids-2017 dataset using machine learning algorithms. In 2022 International Conference on Advances in Computing, Communication and Materials (ICACCM) (pp. 1-10). IEEE.

\bibitem{liu2024elephant}
Liu, Q. and Paparrizos, J., 2024. The elephant in the room: Towards a reliable time-series anomaly detection benchmark. Advances in Neural Information Processing Systems, 37, pp.108231-108261.

\bibitem{ahmed2024explainable}
Ahmed, U., Jiangbin, Z., Almogren, A., Khan, S., Sadiq, M.T., Altameem, A. and Rehman, A.U., 2024. Explainable AI-based innovative hybrid ensemble model for intrusion detection. Journal of Cloud Computing, 13(1), p.150.

\bibitem{sinha2025efficient}
Sinha, P., Sahu, D., Prakash, S., Rathore, R.S., Dixit, P., Pandey, V.K. and Hunko, I., 2025. An efficient data driven framework for intrusion detection in wireless sensor networks using deep learning. Scientific Reports, 15(1), p.34046.

\bibitem{bijlwan2026deep}
Bijlwan, V.P. and Kumar, N., 2026. Deep reinforcement learning in software defined networking: a survey, research challenges and future perspectives. IEEE Communications Surveys and Tutorials.

\bibitem{hadi2026genti}
Hadi, H.J., Khan, M.K., Ahmad, N., Yasmin, R. and Shoker, A., 2026. GenTI: Benchmarking LLMs for Autonomous IDPS Rule Generation for Unseen Attacks. IEEE Transactions on Network Science and Engineering.

\bibitem{owusu2025generalizable}
Owusu, Evans and Rahouti, Mohamed and Verma, Dinesh and Xin, Yufeng and Hsu, D Frank and Schweikert, Christina, 2025.
Generalizable Multi-Model Fusion for Multi-Class DoS Detection Using Cognitive Diversity and Rank-Score Analysis. ACM Transactions on Privacy and Security, 28(3), pp. 1-37.

\bibitem{yakubu2026automated}
Yakubu, Paul Badu and Santana, Lesther and Rahouti, Mohamed and Xin, Yufeng and Chehri, Abdellah and Aledhari, Mohammed, 2026.
Automated and Explainable Denial of Service Analysis for AI-Driven Intrusion Detection Systems. 
IET Information Security, 2026(1),
p. 7264961.

\bibitem{martinez2024redefining}
Martinez, Fernando and Mapkar, Mariyam and Alfatemi, Ali and Rahouti, Mohamed and Xin, Yufeng and Xiong, Kaiqi and Ghani, Nasir, 2024.
Redefining ddos attack detection using a dual-space prototypical network-based approach. 33rd International Conference on Computer Communications and Networks (ICCCN), IEEE, pp.1-9.

\bibitem{wei2023xnids}
Wei, Feng, Hongda Li, Ziming Zhao, and Hongxin Hu, 2023. 
{xNIDS}: Explaining deep learning-based network intrusion detection systems for active intrusion responses." 32nd USENIX Security Symposium (USENIX Security 23), pp. 4337-4354.

\end{thebibliography}
\end{document}